\documentclass[AMA,LATO1COL,demo]{WileyNJD-v2}

\articletype{}

\received{9 January 2017}
\revised{25 October 2017}
\accepted{Pending}

\usepackage{amsmath} 
\usepackage{amsthm}
\usepackage{graphicx} 
\usepackage{bbm}
\usepackage{verbatim}
\usepackage{setspace} 
\usepackage{color}
\usepackage{enumerate}
\usepackage{rotating}
\usepackage{array}
\usepackage{url}
\usepackage{float}
\usepackage{algorithm}
\usepackage{grffile}
\usepackage{lscape}

\usepackage{xcolor}

\newcommand{\bbeta}{ \mbox{\boldmath $\beta $} }

\newcommand{\bY}{\textbf{Y}}
\begin{document}

\title{MEBoost: Variable Selection in the Presence of Measurement Error}

\author[1]{Ben Brown}

\author[2]{Timothy Weaver}

\author[3]{Julian Wolfson}

\authormark{BROWN \textsc{et al}}

\address[1]{\orgname{NAMSA},
\orgaddress{\state{Minnesota}, \country{USA}}}

\address[2]{\orgname{Minneapolis Medical Research Foundation},
\orgaddress{\state{Minnesota}, \country{USA}}}

\address[3]{\orgdiv{Division of Biostatistics}, \orgname{School of Public Health, University of Minnesota},
\orgaddress{\state{Minnesota}, \country{USA}}}

\corres{Ben Brown, 400 US-Hwy 169, Minneapolis, MN 55426 \email{bbrown@namsa.com}}

\presentaddress{400 US-Hwy 169, Minneapolis, MN 55426}

\abstract[Summary]{We present a novel method for variable selection in regression models when covariates are measured with error. The iterative algorithm we propose, Measurement Error Boost (MEBoost), follows a path defined by estimating equations that correct for covariate measurement error. Via simulation, we evaluated our method and compare its performance to the recently-proposed Convex Conditioned Lasso (CoCoLasso) and to the ``naive'' Lasso which does not correct for measurement error. Increasing the degree of measurement error increased prediction error and decreased the probability of accurate covariate selection, but this loss of accuracy was least pronounced when using MEBoost. We illustrate the use of MEBoost in practice by analyzing data from the Box Lunch Study, a clinical trial in nutrition where several variables are based on self-report and hence measured with error.}


\maketitle

\section{Introduction}


Variable selection is a well-studied problem in situations where covariates are measured without error. However, it is common for covariate measurements to be error-prone or subject to random variation around some mean value. Consider, for instance, a study wherein subjects report their daily food intake on the basis of a dietary recall questionnaire. There is variation from day to day in an individual's calorie consumption, but it is also well established in the nutrition literature that there is error associated with the recall or measurement of the number of calories in a meal \cite{Spiegelman1997, Fraser2012}. In the usual regression setting, ignoring measurement error leads to biased coefficient estimation \cite{Rosner1992}, and hence the presence of measurement error has the potential to affect the performance of variable selection procedures. In this example, we may be able to create a predictive model based on these mismeasured dietery recall data, that we can then apply the model to more expensive data that can be measured with reduced or eliminated measurement error such as with the help of a nutritionist or through prepackaged meals.

There has been relatively little research done about variable selection in the presence of measurement error. Sorensen \cite{Sorensen2012} introduced a variation of the Lasso that allows for Normal, i.i.d., additive covariate measurement error. Datta and Zou \cite{Datta2017} proposed the convex conditioned Lasso (CoCoLasso) which corrects for both additive and multiplicative measurement error in the normal case. Both of these methods are applicable to linear models for continuous outcomes, but do not easily extend to regression models for other outcome types (e.g., binary or count data). Meanwhile, there is a  sizable statistical literature on methods for performing estimation and inference for low-dimensional regression parameters in the presence of measurement error \cite{Rosner1992, Stefanski1985, Fuller1987}, but these approaches do not address the variable selection problem and cannot be applied in large $p$, small $n$ problems.




We propose a novel method for variable selection in the presence of measurement error, MEBoost, which leverages estimating equations that have been proposed for low-dimensional estimation and inference in this setting. MEBoost is a computationally efficient path-following algorithm that moves iteratively in directions defined by these estimating equations, only requiring the calculation (not the solution) of an estimating equation at each step. As a result, it is much faster than alternative approaches involving, e.g., a matrix projection calculation at each step. MEBoost is also flexible: the version that we describe is based on estimating equations proposed by Nakamura \cite{Nakamura1990}, which apply to some generalized linear models, and the underlying MEBoost algorithm can easily incorporate measurement error-corrected estimating equations for other regression models.  We conducted a simulation study to compare MEBoost to the Convex Conditioned Lasso (CoCoLasso) proposed by Daata and Zou \cite{Datta2017} and the ``naive'' Lasso which ignores measurement error.  We also applied MEBoost to data from the Box Lunch Study, a clinical trial in nutrition where caloric intake across a number of food categories was based on self-report and hence measured with error.

\section{Background}
\subsection{Regression in the Presence of Covariate Measurement Error}

Our discussion of measurement error models draws heavily from Fuller \cite{Fuller1987}. 
When modeling error the covariates can be treated as random or fixed values. Structural models consider the covariates to be random quantities and functional models consider the covariates to be fixed \cite{Buonaccorsi2010}. We consider a structural model. Let $\mathbf{Y} = \mathbf{X}\beta + \epsilon$, where $\mathbf{X}$ is a (random) matrix of covariates of dimension $n \times p$, $\beta$ a vector of coefficients of length $p$, $\epsilon$ is a vector of Normally distributed i.i.d. random errors of length $n$, and $\mathbf{Y}$ is the resultant outcome vector also of length $n$. In an additive measurement error model, we assume that what is observed is not $\mathbf{X}$ but rather the ``contaminated'' or ``error-prone'' matrix $\mathbf{W = X+U}$ where $\mathbf{U}$ a random $n \times p$ matrix.

When a model is fit that ignores measurement error, i.e. it assumes that the true model is $\mathbf{Y} = \mathbf{W} \mathbf{\beta_{W}} + \epsilon$, the resulting estimates $\mathbf{\hat \beta_{W}}$ are said to be \emph{naive} and satisfy 
\begin{equation}
E[\mathbf{\hat{\beta}_{W}^{'}}] = \mathbf{\beta^{'}}(\mathbf{\Sigma_{XX}}+\Delta)^{-1}\mathbf{\Sigma_{XX}}
\label{eq:attenuation}
\end{equation}
where $\mathbf{\beta}$ is the true coefficient vector, $\mathbf{\Sigma_{XX}}$ is the covariance matrix of the covariates and $\Delta \equiv \Sigma_{UU}$ is the covariance matrix of the measurement error. In the case of linear regression with a single covariate, \eqref{eq:attenuation} simplifies to an attenuating factor that biases the coefficient estimates towards zero. However, with multiple covariates the bias may increase, decrease, and even change the sign of the estimated coefficients. Notably, measurement error affecting a single covariate can bias coefficient estimates in all of the covariates, even those that are not measured with error \cite{Buonaccorsi2010}.

\subsection{Variable selection in the Presence of Measurement Error}

Ma \cite{Ma2010} presented methods to account for measurement error while performing variable selection in parametric and semi-parametric settings. Focusing on the parametric setting, they proposed a wide scoping method that can be used in more than just generalized linear models.  The method relies on deriving the full likelihood of each observation and it's corresponding score function, $S^*_{eff} (\mathbf{W}_i,\bY_i,\beta)$, choosing a penalty function and finding its derivative, $p'(\beta)$, then solving the penalized estimating equations:

\begin{equation}
\sum^n_{i=1} S^*_{eff} (\mathbf{W}_i,\bY_i,\beta)-n p'(\beta) = 0
\label{eq:penme}
\end{equation} 

Solving the penalized equations can be very difficult computationally, especially in the high dimensional setting.  Therefore, we will look to compare our method with faster methods that are variants of the Lasso, which can be solved much more quickly.

\subsection{Lasso in the Presence of Measurement Error}

Sorensen et al. \cite{Sorensen2012} analyze the Lasso \cite{Tibshirani1996} in the presence of measurement error by studying the properties of
\begin{equation}
\hat \beta_{Lasso, \lambda_n} = \operatornamewithlimits{argmin}_{\alpha} \left( ||\mathbf{Y} - \mathbf{W} \alpha||_2^2 + \lambda_n ||\alpha||_1 \right).
\label{eq:lasso}
\end{equation} 

$\hat \beta_{Lasso, \lambda_n}$ is asymptotically biased when $\lambda_{n}/n \rightarrow 0\; as\; n \rightarrow \infty$ since $E[\mathbf{\hat{\beta}^{'}}_{Lasso,\lambda_n}] = \mathbf{\beta^{'}}(\mathbf{\Sigma_{XX}}+\Delta)^{-1}\mathbf{\Sigma_{XX}}$. Notice this is the same bias that is introduced when naive linear regression is performed on observed covariates. 
Sorensen et al. \cite{Sorensen2012} derive a lower bound on the magnitude of the non-zero coefficient elements below which the corresponding covariate will not be selected, and an upper bound on the $L_1$ estimation error $|| \mathbf{\hat \beta_{W}} - \mathbf{\beta}||_1$. They show that with increasing measurement error the lower bound increases, i.e., increasing measurement error adds non-informative noise to the system and so for the signal associated with the relevant covariates to be identified the signal must increase. Increased measurement error also leads to an increase in the upper bound of the estimation error.
Sign consistent selection is also impacted by the presence of covariate measurement error. Sorensen et al. \cite{Sorensen2012} set a lower bound on the probability of sign consistent selection in this setting. The result requires that the \textit{Irrepresentability Condition with Measurement Error} (IC-ME) holds. The IC-ME requires that the measurements of the relevant and irrelevant covariates have limited correlation, relative to the size of the relevant measured covariate correlation. Note the sample correlation of the irrelevant covariates is not considered. By studying the form of the lower bound, it can be concluded that (at least when using the Lasso) measurement error introduces a greater distortion on the selection of irrelevant covariates than it does in the selection of relevant covariates. 

Sorensen et al. \cite{Sorensen2012} introduced an iterative method to obtain the Regularized Corrected Lasso with constraint on the radius R:
\begin{align}
\hat{\beta}_{RCL}= \underset{||\beta||_1 < R}{\text{arg min}} \{||y-W\beta||_2 - n\beta' \Delta \beta + \lambda ||\beta||_1\}.
\end{align}
The main results of their simulation study were consistent with their analytical results, namely that the corrected Lasso had a slightly lower selection rate for the true covariates than the naive Lasso, but was also more conservative in including irrelevant covariates. Further, the prediction error, as measured by both $||\hat{\beta} - \beta||_1$ and $||\hat{\beta} - \beta||_2$, was lower for the corrected Lasso. 

The major drawback of the corrected Lasso method is that it is very computationally intensive, involving an iterative calculation where each step involves a projection of an updated $\hat{\beta}$ onto the $L_1$-ball for a given radius R.  The iterative process must be conducted for each fixed value of the radius R. The selected values of R provide a path of possible solutions for $\hat{\beta}_{RCL}$. Hence, the approach seems impractical for large-scale problems and for repeated application in a simulation study. 

\subsection{The Convex Conditioned Lasso (CoCoLasso)}

A recent paper by Datta and Zou \cite{Datta2017} proposes an alternative approach which they refer to as the Convex Conditioned Lasso (CoCoLasso). Consider the following reformulation of the Lasso problem,
\begin{align}
\hat{\beta_L}(\lambda) &= \underset{\beta}{\text{arg min}} \frac{1}{2}\beta^{'} \Sigma \beta - \rho^{'} \beta + \lambda || \beta||_1. \label{eq:relasso}
\end{align}
The CoCoLasso is based on the Loh and Wainwright corrections \cite{Loh2012} for the predictor-outcome correlation $\rho$ and variance matrix $\Sigma$ in the presence of measurement error. When error-prone covariates $\mathbf{W}$ are measured in place of $\mathbf{X}$, we can get corrected estimates $\tilde \rho$ and $\hat \Sigma$:
\begin{align}
\tilde{\rho} &= \frac{1}{n} \mathbf{W}^{'} \mathbf{Y} \label{eq:coco.rho}&
\hat{\Sigma} &= \mathbf{W}^{'} \mathbf{W} - \Delta
\end{align}
where $\Delta$ is the (assumed known) variance in the measured $\mathbf{W}$. These estimators are unbiased. A measurement error corrected Lasso estimate could then be derived by substituting $\tilde \rho$ and $\hat \Sigma$ into \eqref{eq:relasso}. The problem with this idea is that the corrected matrix $\hat{\Sigma}$ may not be a valid covariance matrix, since it is possible to be non positive semi-definite. If $\hat{\Sigma}$ has a negative eigenvalue, then this Lasso function would be non-convex and unbounded. To overcome this obstacle, the key to the CoCoLasso \cite{Datta2017} is calculating the projection of $\hat{\Sigma}$ onto the space of positive definite matrices:
\begin{align}
(\hat{\Sigma})_{+} = \underset{\Sigma \geq 0}{\text{arg min}} ||\hat{\Sigma} - \Sigma||_{max}. \label{eq:coco.sigma}
\end{align}
The CoCoLasso then solves a standard Lasso problem in which $\hat{\Sigma}$ and $\rho$ with the corrected values from \eqref{eq:coco.rho} and \eqref{eq:coco.sigma}, yielding the CoCoLasso estimator:
\begin{align}
\hat{\beta_C}(\lambda) &= \underset{\beta}{\text{arg min}} \frac{1}{2}\beta^{'} (\hat{\Sigma})_{+} \beta - \tilde{\rho}^{'} \beta + \lambda || \beta||_1.
\end{align}
When $\hat{\Sigma}$ is not positive definite, the projection from \eqref{eq:coco.sigma} can be challenging to compute.  However, the projection only needs to be done once, unlike the Sorensen correction \cite{Sorensen2012} which requires a projection at each iteration.


\section{MEBoost: Measurement Error Boosting}

Our proposed variable selection algorithm, MEBoost (\textbf{M}easurement \textbf{E}rror \textbf{Boost}ing), is based on an iterative functional gradient descent type algorithm that generates variable selection paths. The key idea is that, instead of following a path defined by the gradient of a loss function (e.g., the likelihood), the ``descent'' follows the direction defined by an estimating equation $\mathbf{g(Y,X,\beta)}$. The algorithmic structure of MEBoost is shared with ThrEEBoost (Thresholded Estimating Equation Boost, \cite{Brown2017a}), a general-purpose technique for variable selection based on estimating equations. While ThrEEBoost described an approach to performing variable selection in the presence of correlated outcomes by leveraging the Generalized Estimating Equations \cite{Liang1986}, MEBoost achieves improved variable selection performance in the presence of measurement error by following a path defined by a measurement error corrected score function due to Nakamura which is described in Section \ref{sec:corrscore}.  Nakamura's approach is applicable to linear regression models with normal additive or multiplicative measurement error. Closed-form corrected score functions are also derived for Poisson, Gamma, and Wald regression.  Nakamura comments that no closed form correction can be created for logistic regression. By using this family of corrected score functions, the MEBoost algorithm is more broadly applicable than the corrected Lasso and CoCoLasso, neither of which is obviously generalizable beyond linear regression.

\subsection{Corrected Score Function}
\label{sec:corrscore}

Nakamura \cite{Nakamura1990} proposed a set of corrected score functions for performing estimation and inference in the generalized linear regression model where covariates are subject to additive measurement error with known variance matrix $\Delta$. In general, the corrected score function S* based on the covariates measured with error ($\mathbf{W}$), has the expectation equal to the score function, S, based on the true covariates ($\mathbf{X}$). For the normal linear model,
Nakamura proposed the following correction to the negative log-likelihood to account for measurement error:
\begin{align}
l^{*}(\mathbf{Y},\mathbf{W},\beta)^{'} =
-\frac{n}{2}log(2\pi)-n log(\sigma)-\frac{1}{2\sigma^2} \sum [(y_i-\beta'w_i)^2-\beta'
\Delta\beta] \label{eq:corrlike}
\end{align}

Differentiating \eqref{eq:corrlike} with respect to $\beta$, we obtain the corrected score function:
\begin{align}
S^{*}(\mathbf{Y},\mathbf{W},\beta)^{'} = S(\mathbf{Y},\mathbf{W},\beta)^{'}+n\sigma^{-2}\beta^{'}\Delta.
\end{align}

In this case the corrected score function is the 'naive' score function, $S(\mathbf{Y},\mathbf{W},\beta)^{'}$, with a measurement error correction determined by the sample size, model error, measurement errors, and the coefficient value: $n\sigma^{-2}\beta^{'}\Delta$. The naive score function is the score function from the true model calculated with the measured covariates:
\begin{align}
S(\mathbf{Y},\mathbf{W},\beta,\sigma) = \sigma^{-2}\left(\mathbf{W}^{'}\mathbf{Y}-\mathbf{W}^{'}\mathbf{W}\beta\right)
\end{align}
The corrected variance estimate will be calculated as $\partial l^{*}/\partial \sigma = 0$, which in the normal case is:
\begin{align}
\hat{\sigma}^2 = n^{-1}\left(\mathbf{Y} - \mathbf{W}\beta^{*}\right)^{'}\left(\mathbf{Y} - \mathbf{W}\beta^{*}\right) - \beta^{*'}\Delta\beta^{*}.
\end{align}
Similarly to the corrected score function, the corrected variance estimate is the naive variance estimate, $n^{-1}\left(\mathbf{Y} - \mathbf{W}\beta^{*}\right)^{'}\left(\mathbf{Y} - \mathbf{W}\beta^{*}\right)$, with a measurement error correction. The correction reduces the estimated variance, thus subtracting the noise introduced by the measurement error. In the variance case the correction factor is determined only by the true coefficient vector and the measurement error variance. 

As another example, the correction for Poisson distributed data is the following:
\begin{align}
S(\mathbf{Y},\mathbf{W},\beta) = \sum [y_k w_k -(w_k - \Delta\beta) exp(\beta'w_k- \beta'\Delta\beta/2)]
\end{align}
which we apply in our data application (see Section \ref{sec:application}). Nakamura \cite{Nakamura1990} also provides corrections for multiplicative measurement error in linear regression, as well as measurement error in Gamma and Wald regression.  In what follows, we use the normal linear additive measurement error corrected score function as part of an iterative path-following algorithm that performs variable selection in the presence of covariate measurement error.

\subsection{The MEBoost Algorithm}


Our proposed variable selection algorithm, MEBoost, consists of applying ThrEEBoost with the corrected score function and corrected variance estimate described in the previous section. Algorithm \ref{algo:meboost} summarizes the MEBoost procedure.


Let $\tau \in [0,1]$ be the fixed thresholding parameter. Starting with a $\beta$ estimate of \textbf{0} and a $\hat{\sigma}^{2} = 1$, the corrected score function $\mathbf{S^*}$ is calculated at these values, and the magnitude of each component of $\nu \equiv \mathbf{S^*}$ is recorded. The indices of elements to update are identified by a thresholding rule, $J_t = \{ j: |\nu_j| \geq \tau \cdot \text{max}_j |\nu_j| \}$. The next point in the variable selection path, $\beta^{(1)}$, is obtained by adding a small value, $\gamma$, to each of these elements in the direction corresponding to the signs of each $\nu_j$ for $j \in J_t$. This updated $\beta^{(1)}$ is used to calculate an updated corrected $\sigma^{2(1)}$. The algorithm continues for $T$ iterations, where $T$ is typically chosen to be large (e.g., 1,000). 

\begin{algorithm}
	\caption{MEBoost}
	\label{algo:meboost}
	\begin{algorithmic}
		\Procedure{MEBoost}{}\\
		Set $\bbeta^{(0)}$ = {\bf 0}\\
		Set $\sigma^{2,(t=0)} = 1$
		\For{$t = 0, \dots, T$}
		\State Compute $ \nu = \mathbf{S^*}(\mathbf{Y},\mathbf{W}, \bbeta)_{\bbeta=\bbeta^{(t-1)}}$
		\State Identify $J_t = \{ j: |\nu_j| \geq \tau \cdot \text{max}_j |\nu_j| \}$
		\For{ all $j_t \in J_t$}
		\State Update $\bbeta_{j_t}^{(t)} = \bbeta_{j_t}^{(t-1)} + \gamma $ sign($\nu_{j_t}$)
		\EndFor
		\State Set $\sigma^{2,(t)} = n^{-1}\left(\mathbf{Y} - \mathbf{W}\beta^{(t)}\right)^{'}\left(\mathbf{Y} - \mathbf{W}\beta^{(t)}\right) - \beta^{(t)'}\Delta\beta^{(t)}$
		\EndFor
		\EndProcedure
	\end{algorithmic}
\end{algorithm}

The parameters $\gamma$, $T$ and $\tau$ interact to determine the specific variable selection path that results from the algorithm. The smaller the value of $\gamma$ the smaller the distance between $\beta$ estimates on the selection path, while a larger value of $\gamma$ leads to larger jumps in the selection path. Ideally, a very small value of $\gamma$ (e.g., 0.01). would be used, but if $||\beta||_{\mathbf{1}}$ is large, a large number of iterations, $T$, may be required to generate a selection path. This of course is the trade-off one is required to make when determining the step size. A selection path incremented by only a small value is preferable to a path which takes large steps, but the time required for a large number of iterations may become prohibitive. With each of the $t$ iterations those elements of the coefficient vector that are still of size zero have not been selected at this iteration. A conservative selection approach takes a combination of small $\gamma$ and $T$, whereas a more aggressive approach takes a combination of larger value $\gamma$ and $T$. In the case when $\tau = 1$, the MEBoost algorithm only updates the element(s) with the maximum absolute value. For any combination of $\gamma$ and $T$, this is the most conservative approach that can be taken and will lead to sparser models than when a threshold is considered. It also requires a much larger value of $T$. 

The parameter $\tau$ determines how many coefficients are updated at each iteration; it offers a compromise between updating each coefficient at every iteration ($\tau = 0$, similar to standard gradient descent) and updating only the coefficient corresponding to the element of the estimating equation with largest magnitude ($\tau = 1$). In the context of Generalized Linear Models without measurement error, Wolfson \cite{Wolfson2011} showed that setting $\tau = 1$ yields an update rule that is asymptotically equivalent (as $T \rightarrow \infty$, $\gamma \rightarrow 0$, and $T \cdot \gamma \rightarrow 0$) to following the path of minimizers of an $L_1$-penalized projected artificial log-likelihood ratio whose tangent is the GLM score function. 
In the case when $\tau = 1$, the MEBoost algorithm only updates the element(s) with the maximum absolute value. For any combination of $\gamma$ and $T$, this is the most conservative approach that can be taken and will lead to sparser models than when a threshold is considered. It also requires a much larger value of $T$. 
By allowing multiple directions to be updated at each iteration, MEBoost can explore a much wider range of variable selection paths; as we discuss later, cross-validation can be used to select the parameter $\tau$ which leads to the optimal level of thresholding. In ThrEEBoost \cite{Brown2017a}, it was shown that a threshold in the range of 0.4-0.8 may perform better than thresholds closer to 0 or 1.

\subsubsection{Connection to the ME-Lasso}

Nakamura's measurement corrected score functions are derivatives of corrected negative log likelihoods. In the normal case, the correction is exactly that described in Sorensen et al. (see Equation \eqref{eq:lasso}). Hence, the arguments of Rosset \cite{Rosset2004} can be applied to show that 1) MEBoost applied with $S^*$ and threshold value $\tau = 1$, and 2) the solutions to \eqref{eq:lasso}, have the same local behavior. Specifically, under some regularity conditions \cite{Wolfson2011}, as $T \rightarrow \infty$ and $\epsilon \rightarrow 0$ with $T \cdot \epsilon \rightarrow 0$, MEBoost's iterative steps match the sequence of solutions to \eqref{eq:lasso}.

\subsubsection{Selecting a final model}

For a fixed $\tau$, identifying a final model involves choosing a point on the variable selection path generated by Algorithm \ref{algo:meboost}; this is akin to choosing the penalty parameter in the Lasso. Cross-validation using a loss function relevant to the problem at hand (e.g., mean squared error) can be used to select a $\hat{\beta}$ on the path. Cross-validation can similarly be used to select the best value of $\tau$. The full procedure is described in Algorithm \ref{algo:sel}. 

\begin{algorithm}
	\caption{Model Selection for MEBoost}
	\label{algo:sel}
	\begin{algorithmic}
		\Procedure{Cross Validation}{}
		\State Divide the observations into $K$ folds where $\frac{1}{K}$ of the observations are used as a test set.
		\For{$k = 1, \dots, K$}
		\State Apply MEBoost for fixed value $\tau$.
		\State Obtain the mean squared prediction error of each candidate model on the test set.
		\State Calculate $||\hat{\beta}||_1^{(k)}$ for the $\hat{\beta}^{(t)}$ that minimizes mean squared prediction error. 
		\EndFor
		\State Repeat across the $K$ possible test sets and compute the mean of the selected $||\hat{\beta}||_1^{(k)}$'s.
		\EndProcedure
	\end{algorithmic}
\end{algorithm}

\section{Simulation Study}

To examine the impact of measurement error in the covariates on variable selection we performed a simulation study. We evaluated MEBoost by comparing it to two variable selection methods: the Convex Conditioned Lasso (CoCoLasso), and the ``naive'' Lasso which does not correct for measurement error.

\subsection{Simulation Set-up}

Data were generated from a linear regression model with iid normal errors,
$\mathbf{Y} = \mathbf{X} \mathbf{\beta} + \mathbf{\epsilon}$; where $\mathbf{\epsilon_i} \sim N(\mathbf{0}, \sigma^2_\epsilon)$ and $\sigma_\epsilon = 1.5$. The sample size for all studies is 80. The true covariates are drawn from a multivariate normal distribution, $\mathbf{X} \sim MVN(\mathbf{0},\Sigma_{XX})$. $\Sigma_{XX}$ is a block diagonal matrix with diagonal entries equal to 1, and 10 by 10 blocks corresponding to a group of 10 covariates with an exchangeable correlation structure with common pairwise correlation $\phi=0.3$. In all simulations the true model has 10 non-zero coefficients and 90 zero coefficients, i.e., $\beta = (\mathbf{1}_{10},\mathbf{0}_{90})$, so that the relevant covariates in the first block were correlated. 


The measured covariates were generated as $\mathbf{W} = \mathbf{X} + \mathbf{U}$ for $\mathbf{U}$ a matrix whose columns were generated as described below. To explore the impact of different types of measurement error, we considered 10 different scenarios for generating the columns of $U$ and varying the assumptions made about it. In the first five scenarios, $\mathbf{U}$ is assumed to be normally distributed with mean zero and covariance matrix $\Omega$, and the scenarios explore different structures for $\Omega$. In each of Scenarios 1-5, we correctly specify the distribution of $\mathbf{U}$ when applying MEBoost and the CoCoLasso. Scenarios 6-10 explore cases where the distribution of $\mathbf{U}$ is incorrectly specified.


\begin{enumerate}
\item Base case: $\mathbf{U} \sim N( \mathbf{0}, \delta^2 \Omega_1)$, where $\Omega_1 = \mathbf{I}$ the identity matrix, and $\delta^2 = 0.75$.

\item Varying $\delta^2$:  $\delta_j^2=0.3375+0.075j$ for j in 1-10. This pattern repeats across the blocks of 10 covariates. The relevant covariates have similar spreads of measurement error to the irrelevant covariates. 

\item Correlated Measurement Error: Measurement error was assumed normally distributed with an exchangeable correlation structure Within each block, $\mathbf{U} \sim N( \mathbf{0}, \delta^2 \Omega_3 )$ where $\Omega_3 = \rho \mathbf{1}\mathbf{1'} + (1-\rho) \mathbf{I}$, $\delta^2 = 0.75$, $\rho = 0.3$, and $\mathbf{1}$ is vector of ones. 

\item Varying $\delta^2$ with correlation: $\mathbf{U}$ is distributed normally and is centered at 0 with $\delta_j^2=0.3375+0.075j$ for j in 1-10. This pattern repeats across the blocks of 10 covariates.  The relevant covariates have similar spreads of measurement error to the irrelevant covariates. In blocks of ten, there is correlation in $\mathbf{U}$ of 0.3

\item Some $\mathbf{U}$'s not measured with error: $\mathbf{U} \sim N(0, \delta^2 \Omega_5)$, where $\delta^2=0.75$, $diag(\Omega_5) = [0,  1, 0, 1, \dots]$ and $\Omega_{5,ij} = 0$ for $i \neq j$. 

\item Overestimated $\delta^2$: $\mathbf{U}$ generated as in Scenario 1, but we specify $\delta^2=1.5$.

\item Underestimated $\delta^2$: $\mathbf{U}$ generated as in Scenario 1, but we specify $\delta^2=0.375$.

\item Misspecified correlation: $\mathbf{U}$ generated as in Scenario 3, but we ignore the correlation and specify $\Omega = \delta^2 \mathbf{I}$ in running MEBoost and CoCoLasso. 

\item Measurement error is distributed uniformly: Each entry $\mathbf{U}_{ij}$ of $\mathbf{U}$ is generated independently from a Uniform distribution, $\mathbf{U}_{ij} \sim U(-1.5, 1.5)$. MEBoost and CoCoLasso are run assuming $\mathbf{U} \sim N(\mathbf{0}, \delta^2 \mathbf{I})$ with $\delta^2 = 0.75 = Var(\mathbf{U}_{ij})$. 

\item Measurement error is distributed asymmetrically: Each entry $\mathbf{U}_{ij}$ of $\mathbf{U}$ is generated independently from a shifted exponential distribution, $\mathbf{U}_{ij} + \sqrt{0.75} \sim exp(\sqrt{0.75})$. MEBoost and CoCoLasso are run assuming $\mathbf{U} \sim N(\mathbf{0}, \delta^2 \mathbf{I})$ with $\delta^2 = 0.75 = Var(\mathbf{U}_{ij})$. 

\end{enumerate}


MEBoost was performed for each threshold value in the set $\tau=(0.0,0.2,0.4,0.6,0.8,1.0)$, and cross-validation (using the error-prone covariates) was used to select the optimal value of $\tau$ and number of MEBoost iterations, as well as the value of $\lambda$ in the CoCoLasso and naive Lasso. We compared MEBoost, CoCoLasso, and naive Lasso on two metrics of prediction error: mean squared error based on the true covariates (MSE = $\frac{1}{n}(\mathbf{Y} - \mathbf{X}\hat{\beta})^{'}(\mathbf{Y} - \mathbf{X}\hat{\beta})$), mean squared error prediction based on the measured covariates (MSE-M = $\frac{1}{n}(\mathbf{Y} - \mathbf{W}\hat{\beta})^{'}(\mathbf{Y} - \mathbf{W}\hat{\beta})$). These metrics were estimated using independent test sets generated during each individual simulation. We also computed $L_1$ distance from the true $\beta$, and variable selection sensitivity and specificity. For each scenario the metrics presented are the average over 1,000 simulations, and are calculated at intervals of 0.05 along $||\hat{\beta}||_1 \in \left\{0.05,0.1,0.15,...,15\right\}$; the true value, $||\beta||_1 = 10$. Because the MEBoost algorithm may change multiple indices at each iteration it may not have values along each interval in the path. To account for this, a linear approximation of the relevant statistic was made at each point in the path.

We note that in this simulation study we chose to investigate model performance based on both the true and error-prone covariates. The motivation for techniques like ours which account for measurement error is to uncover the underlying relationship between the error-free covariates $\mathbf{X}$ and the outcome $\mathbf{Y}$. Hence, in an ideal world, values of $\mathbf{X}$ would be available on some subset (or an independent set) of observations so that prediction error could be assessed and the ``best'' model chosen. However, in practice we will often only have access to the error-prone covariates $\mathbf{W}$ for model fitting. So, if error-free measurements $\mathbf{X}$ are not (and may never be) available, is it worthwhile to correct for measurement error?  Buonaccorsi \cite{Buonaccorsi1995} argued against correction, using the logic that the future predictions will be based on (error-prone) $\mathbf{W}$, not on (error-free) $\mathbf{X}$. Indeed, it can be shown in simple linear regression, that without the correction in a large sample the expected value of MSE-M is less than or equal to that of an estimate ignoring measurement error. However, as seen in the results section that follows, we found that correcting for measurement error decreased prediction error regardless of whether predictions were computed using error-free or error-prone covariates. Since we often only have mismeasured data available, it is reassuring to see that we are able to use the measured covariates to perform cross-validation to select a model that will provide us with an accurate relationship between the outcome and true covariate. This finding is discussed in greater detail below.



\subsection{Simulation Results}

Table \ref{tab:corresults} presents the minimum MSE, MSE-M, $L_1$ distance from the true $\beta$, sensitivity, and specificity at the minimum MSE for the three variable selection methods across the 10 scenarios.  In all ten scenarios, MEBoost had the lowest MSE, MSE-M, and $L_1$ distance from the true $\beta$.  The CoCoLasso has 16.6\%-71.7\% higher prediction error from the true covariates than MEBoost and in the case where measurement error is overestimated, the prediction error from the CoCoLasso is 5.26 times that of MEBoost. This is likely due to the fact that the Loh and Wainwright correction $\hat \Sigma$ in \eqref{eq:coco.rho} is more negative, and hence requires a ``longer'' (and hence potentially more distorting) projection onto the space of positive definite matrices. 

In terms of variable selection, MEBoost had a greater sensitivity and lower specificity than CoCoLasso in each case while Lasso had the lowest specificity.  The Lasso struggles most when correlation is present in the measurement error.  The MSE is about 2.5 times that of MEBoost, when we allow MEBoost to account for the correlation.  All methods perform poorly when we misspecify $\Delta$ by ignoring the correlation. The sensitivity and specificity are at high levels for most simulations with the exception of the misspecified $\Delta$ that ignored correlation. Overestimating $\delta$ lead to a more conservative selection process with a high specificity, while underestimating $\delta$ had a higher sensitivity. The $L_1$ distance from the true $\beta$ can also tell us about performance.  Again, the scenario where we misspecify $\Delta$ by ignoring correlation performs worst.

\section{Data Application}
\label{sec:application}

We applied our method to baseline data collected in the Box Lunch Study, a randomized trial of the effects of portion size availability on weight change. In the study, a total of 219 subjects were randomized to one of four groups: in three groups, subjects were provided a free daily lunch with a fixed number of calories (400, 800, and 1600).  The control group was not provided a free lunch.  

We considered the problem of predicting the number of times subjects reported binging on food in the last month, using Poisson regression with 99 explanatory variables. All variables were measured at baseline. 16 of the 99 explanatory variables were self-reported measures; of these 16, 8 were measures of food consumption and therefore possibly subject to substantial measurement error we will notate $\delta^2_D$.  Another 8 may have also been measured with error, notated $\delta^2_M$.  Kipnis \cite{Kipnis2003} examined a nutritional study with a 24 hour recall, and found that the correlation between the true and reported consumption of protein and energy was only 0.336.  We assume this relationship exists in each of our variables measured with error.  Assuming the measurement error variance $Var(U_i) \equiv \delta^2_i$ is independent of the variance of the true covariate $Var(X_i) \equiv \sigma^2_{X_i}$, we can obtain:
\begin{equation}
\rho_{W_i, X_i} = \rho_{X_i+U_i,X_i}
=\frac{\sigma^2_{X_i}}{\sqrt{\sigma^2_{X_i}(\sigma^2_{X_i}+ \delta_i^2)}} \ 
\Longrightarrow \ Var(W_i) = 1-\rho_{X_i+U_i,X_i}^2=\frac{\delta_i^2}{\sigma^2_{X_i}+\delta_i^2}
\end{equation}
and hence $Var(W_i)=1-0.336^2=0.887.$ This is the value we will need to provide MEBoost for our assumption of the measurement error.  We assume this level of measurement error for each 24 hour dietary recall variable. After scaling our predictors to have zero mean and unit variance, we applied our method with the Nakamura correction. Since our measured data has its variance ($\delta^2_i+\sigma^2_{X_i}$) scaled to equal 1, we assumed that the 8 dietary recall covariates measured with error had $\hat{\delta}_D^2=0.887$.  Since dietary variables may be more prone to measurement error than other variables, we scaled the assumed error of the other 8 variables to be half that of the nutritional variables: $\hat{\delta}_M^2=\hat{\delta}_D^2/2$. The remaining variables were assumed to be measured without error. We conducted a sensitivity analysis to assess the performance of our method by setting $\hat{\delta}_D^2 = 0.5$ and 0.25. 

To select tuning parameters, we employed 8-fold cross validation based on the deviance on a training set consisting of 70\% of the data.  The performance of our model was evaluated on the remaining test set. We present the models derived from MEBoost performed with three different thresholds $\tau$: 0.2, 0.6 (the approximate value estimated using cross-validation), and 0.9. 

Table \ref{tab:BLScoef} shows the selected variables and estimated prediction error (MSE-M, bottom row) for various MEBoost models along with results from the naive Lasso. We did not compare to the Measurement Error Lasso or the CoCoLasso because implementing these techniques in a problem of this size was computationally infeasible.  The deviance and MSE-M were lowest for the model selected by MEBoost assuming the highest measurement error (= 0.887) and a threshold value of 0.6. This  model ($\hat{\delta}_D^2=0.887$ and $\tau=0.6$) selected just 4 variables, which were a subset of the 7 chosen with the naive Lasso. The other two MEBoost models included up to two additional variables to the MEBoost model that minimized MSE-M (selected with $\hat{\delta}_D^2=0.887$ and $\tau=0.6$). Regardless of the assumption about the level of measurement error, using a threshold value of $\tau=0.2$ leads to the inclusion of several variables with small coefficients, and a much higher deviance and prediction error. Of particular note is that the naive Lasso (and MEBoost with the lower threshold) included the variable corresponding to the number of daily calories consumed at breakfast, while the best-performing MEBoost models (with $\tau = 0.6$ and 0.9) did not. Since it is based on a 24-hour dietary recall, this variable may be particularly susceptible to measurement error induced by recall bias.

\section{Discussion}

We examined the variable selection problem in regression when the number of potential covariates is large compared to the sample size and when these potential covariates are measured with measurement error. We proposed MEBoost, a computationally simple descent-based approach which follows a path determined by measurement error-corrected estimating equations. We compared MEBoost, via simulation and in a real data example, with the recently-proposed Convex Conditioned Lasso (CoCoLasso) as well as the naive Lasso which assumes that covariates are measured without error. In almost all simulation scenarios, MEBoost performed best in terms of prediction error and coefficient bias. The CoCoLasso is more conservative with the highest specificity in each case, but sensitivity and prediction are better with MEBoost. In the comparison of selection paths, we saw that MEBoost was more aggressive in identifying variables to be included in the model more quickly than the CoCoLasso. These differences were most apparent when the measurement error had a larger variance and a more complex correlation structure. Specifically, when faced with a data set of 1000 observations and 1000 covariates, MEBoost obtained a solution in 1.3 seconds, while the CoCoLasso needed 6:17.

As shown in the simulation study, MEBoost has lower prediction error than the Lasso on independent test data when predictions are based on the true (i.e., non-error-prone) covariates. It is interesting to note that MEBoost retains some advantage, albeit a more modest one, over the Lasso when predictions are based on error-prone covariates. This finding appears to contradict the intuition that accounting for covariate measurement error provides no benefit when the goal is prediction and error-free covariates will never be available. However, the observed benefit in our simulation is likely due to the fact that MEBoost is somewhat more flexible than the Lasso as it uses an additional parameter, the threshold $\tau$, which allows it to explore the model space more comprehensively.  Nevertheless, it is reassuring that by using the error-prone covariates to perform cross-validation and select a model, MEBoost still allows us to select a model that offers an improvement in prediction in the setting where we will have correctly measured covariates.



MEBoost, while a promising approach, has some limitations. One limitation--which is shared with many methods that correct for measurement error--is that we assume that the covariance matrix of the measurement error process is known, an assumption which in many settings may be unrealistic. In some cases, it may be possible to estimate these structures using external data sources, but absent such  data one could perform a sensitivity analysis with different measurement error variances and correlation structures, as we demonstrate in the real data application. Another challenging aspect of model selection with error-prone covariates is that, even if the set of candidate models is generated via a technique which accounts for measurement error, the process of selecting a final model (e.g., via cross-validation) still uses covariates that are measured with error. However, we showed in our simulation study that MEBoost performs well in selecting a model which recovers the relationship between the true (error-free) covariates and the outcome, even when using error-prone covariates to select the final model. This finding suggests that the procedure for generating a ``path'' of candidate models has a greater influence on prediction error and variable selection accuracy than the procedure picking a final model from among those candidates.

To conclude, we note that while we only considered linear and Poisson regression in this paper, MEBoost can easily be applied to other regression models by, e.g., using the estimating equations presented by Nakamura \cite{Nakamura1990} or others which correct for measurement error. In contrast, the approaches of Sorensen \cite{Sorensen2012} and Datta \cite{Datta2017} exploit the structure of the linear regression model and it is not obvious how they could be extended to the broader family of generalized linear models. The robustness and simplicity of MEBoost, along with its strong performance against other methods in the linear model case suggests that this novel method is a reliable way to deal with variable selection in the presence of measurement error.

\bibliographystyle{WileyNJD-AMA}
\bibliography{library2}

\clearpage

\section{Tables}

\begin{table}[ht]
\centering
\begin{tabular}{llrrrrr}
 Scenario & Method & MSE & MSE-M & $L_1$D & SENS & SPEC \\ \hline
Measurement error &  MEBoost & 4.86 & 10.65  & 5.17 & 0.95 & 0.86 \\ 
iid & Lasso & 7.13 & 11.63  & 6.75 & 0.98 & 0.75 \\ 
 & CoCoLasso & 6.30 & 12.53 &  6.04 & 0.92 & 0.91 \\ \hline
Varying $\delta$ & MEBoost &  4.88 & 10.52 & 5.21 & 0.96 & 0.85 \\ 
 & Lasso    &  7.08 & 11.42 & 6.76 & 0.98 & 0.76 \\ 
   & CoCoLasso    &  7.18 & 14.84 & 6.50 & 0.85 & 0.95 \\ \hline
Some $\delta=0$ & MEBoost  &  3.65 & 6.70 & 3.57 & 0.99 & 0.87 \\ 
 & Lasso    & 4.88 & 6.23 & 5.42 & 0.99 & 0.83 \\ 
   & CoCoLasso    &  6.23 & 11.15 & 5.60 & 0.92 & 0.95 \\ \hline
Varying $\delta$ & MEBoost  &  6.19 & 19.12 & 6.27 & 0.95 & 0.87 \\ 
 \& correlation & Lasso    &  15.18 & 21.42 & 9.63 & 0.80 & 0.79 \\ 
   & CoCoLasso    & 8.94 & 22.47 & 7.77 & 0.78 & 0.94 \\\hline 
Correlation in  & MEBoost  &  6.16 & 19.27 & 6.29 & 0.95 & 0.87 \\ 
measurement error & Lasso    &  15.78 & 22.24 & 9.75 & 0.80 & 0.79 \\ 
   & CoCoLasso    &  8.67 & 22.29 & 7.81 & 0.78 & 0.94 \\ \hline
Overestimated $\delta$ & MEBoost   & 4.00 & 10.21 & 3.46 & 0.94 & 0.94 \\ 
  & Lasso    &  7.18 & 11.71 & 6.75 & 0.98 & 0.76 \\ 
   & CoCoLasso    &  21.05 & 28.22 & 9.11 & 0.37 & 1.00 \\ \hline
Underestimated $\delta$ & MEBoost  &  5.54 & 10.81 & 6.28 & 0.98 & 0.76 \\ 
  & Lasso   &  7.18 & 11.71 & 6.75 & 0.98 & 0.76 \\ 
  & CoCoLasso    & 6.46 & 12.02 & 6.20 & 0.95 & 0.87 \\ \hline
  Misspecified $\Delta$, &MEBoost  &  12.79 & 21.03 & 9.41 & 0.86 & 0.80 \\ 
ignores correlation  & Lasso    &  15.78 & 22.24 & 9.75 & 0.80 & 0.79 \\ 
  & CoCoLasso    &  15.67 & 24.17 & 9.51 & 0.55 & 0.93 \\ \hline
Measurement error & MEBoost  &  4.89 & 10.68 & 5.16 & 0.95 & 0.85 \\ 
 from asymmetric & Lasso    &  7.21 & 11.85 & 6.81 & 0.98 & 0.75 \\ 
distribution  & CoCoLasso    &  7.32 & 15.20 & 6.71 & 0.85 & 0.95 \\ \hline
Measurement error & MEBoost  &  4.81 & 10.52 & 5.17 & 0.96 & 0.84 \\ 
from uniform  & Lasso    &  7.19 & 11.75 & 6.80 & 0.99 & 0.75 \\ 
distribution  & CoCoLasso    &  6.61 & 13.91 & 6.26 & 0.87 & 0.94 \\ 
\end{tabular}
 \caption{Performance metrics for the 1,000 simulations in various measurement error scenarios. The models were selected at the point with minimum MSE-M. \label{tab:corresults} }
\end{table}



\begin{landscape}
\begin{table}[ht]
\footnotesize
\centering
\begin{tabular}{lcccccccccc}
   & \multicolumn{3}{c}{$\hat{\delta}_D^2=0.887$} &
  \multicolumn{3}{c}{$\hat{\delta}_D^2=0.5$} &
  \multicolumn{3}{c}{$\hat{\delta}_D^2=0.25$} & Naive\\
Variable & $\tau=0.2$ & $\tau=0.6$ & $\tau=0.9$ & $\tau=0.2$ & $\tau=0.6$ & $\tau=0.9$ &$\tau=0.2$ & $\tau=0.6$ & $\tau=0.9$ & Lasso \\ 
  \hline
Ate lg amt past 28 days & 0.09 & 0.23 & 0.39 & 0.08 & 0.23 & 0.33 & 0.09 & 0.22 & 0.29 & 0.21 \\ 
  Lost control past 28 days & 0.09 & 0.23 & 0.21 & 0.08 & 0.23 & 0.22 & 0.09 & 0.22 & 0.23 & 0.24 \\ 
  TFEQ Disinhibition & 0.09 & 0.23 & 0.09 & 0.08 & 0.23 & 0.18 & 0.09 & 0.22 & 0.21 & 0.17 \\ 
  BCT: Max clicks for pizza slice & 0.09 & 0.21 & 0.1 & 0.08 & 0.23 & 0.19 & 0.09 & 0.22 & 0.2 & 0.16 \\ 
  Long fast (0=no days 6=every day) & 0.09 & - & - & 0.08 & - & - & 0.09 & - & - & - \\ 
  Judge your shape (0=not at all 6=markedly) & 0.07 & - & - & 0.07 & 0.06 & - & 0.07 & 0.06 & - & - \\ 
  Judge your weight (0=not at all 6=markedly) & 0.07 & - & - & 0.07 & - & 0.06 & 0.07 & 0.06 & 0.06 & 0.06 \\ 
  Dissatisfied with shape (0=not at all 6=markedly) & 0.07 & - & - & 0.07 & - & - & 0.07 & - & - & - \\ 
  BCT: Rounds of clicking for pizza & 0.06 & - & - & 0.06 & - & - & 0.07 & - & - & - \\ 
  BCT: Pmaxpizza/(Pmaxpizza+Pmaxread) & 0.06 & - & - & 0.06 & - & - & 0.06 & - & - & - \\ 
  TFEQ Hunger & 0.06 & - & - & 0.06 & - & - & 0.06 & - & - & - \\ 
  Dissatisfied with weight (0=not at all 6=markedly) & 0.06 & - & - & 0.06 & - & - & 0.06 & - & - & - \\ 
  NDSR breakfast kcals at BL & 0.06 & - & - & 0.05 & - & - & - & - & - & 0.08 \\ 
  CDRS body image (1=thinnest 9=fattest) & 0.05 & - & - & 0.05 & - & - & 0.05 & - & - & - \\ 
  Dem9 Household income & - & - & - & - & - & - & - & - & - & -0.05 \\ 
  Eat lunch in cafeteria & - & - & - & - & - & - & -0.05 & - & - & - \\ 
  Eat items from home, days/wk & -0.06 & - & - & -0.06 & - & - & -0.06 & - & - & - \\ 
  Cohort first lunch date & -0.07 & - & - & -0.07 & - & - & -0.08 & - & - & - \\ 
   \hline\hline
Deviance & 210.43 & 89.06 & 100.49 & 222.07 & 90.22 & 93.23 & 208.07 & 90.63 & 90.38 & 97.12 \\ 
MSE-M & 22.35 & 5.61 & 8.10 & 23.28 & 5.69 & 7.29 & 22.04 & 5.97 & 6.21 & 7.89 \\\hline
\end{tabular}
 \caption{Coefficients, Deviance, and MSE-M from selected models for MEBoost with specified value of $\tau$ and $\hat{\delta}_D^2$ and the Lasso. Small coefficients (magnitude $<$ 0.05) are omitted. ``-" indicates that the variable was not selected in the model.
 \label{tab:BLScoef} }
\end{table}
\end{landscape}

\end{document}